# USAGE OF MECHANICALLY SWITCHING DEVICES FOR HV ELECTROSTATIC ELEMENTS OF BEAM OPTICS

Alexander Mikhailichenko, CHESS, Ithaca, NY 14853

*Abstract.* We are considering usage of gas-filled mechanical relays- Kilovacs and Gigavacs-for power supply of HV electrical Dipoles, Quadrupoles *etc*. where even a time dependent regime required. One can expect the physical switching time on few nanosecond level with such devices operating up to 70 *kV* with currents of ~50 *A*.

## 1. INTRODUCTION

Mostly elements of beam focusing systems used in storage rings and beam transport channels are the magnetic ones: Dipoles, Quadrupoles *etc*. Electrostatic elements are represented there as some specific elements such as kickers, electrostatic separators, septa, quadrupoles *etc*. For example, usage of electrostatic quadrupoles for vertical focusing of muons in dedicated (*g-2*) ring is a key concept of E-989 (E-821) experiment performed for measurement of anomalous magnetic moment of muon. One peculiarity with quadrupoles in this experiment is that the dipole component of electric field should be introduced for a short period of time of the order 16*μs* for scrapping particles with large amplitudes [1].

Some minor comments could be done about technical realization of electric scheme in E-821. For example the HV feedthroughs should be not attached to the flange by tubes (see Fig.8 of [1])-as the HV wires should be a shadow of insulator. So the necks of these wires at the level of flange plane should not be exposed to the elevated field strength allowing development of corona discharge.

In [2] described the scheme for Quadrupole PS for E-989 experiment. Pretty much it is the same scheme as it is described in [1]. One statement could be criticized there too. The maximal current in a circuit which runs for a sub microsecond level associated with charging the parasitic capacitance associated with feeding cables ($C_4=C_6=3nF$, Fig.1 [2]). The maximal level of current comes to be $I_{max}=U/R_6=32kV/100\Omega=320A$, which lasts for a time $\tau=R_6C_4=100\Omega \times 3nF=300ns$ only. But the authors come to conclusion that the PS should be able to generate this ~400A current for entire millisecond, suggesting thyratron switches for T1-T3. One obvious solution for reduction of this current (and type of switch) associated with reduction of parasitic capacitance and increase of serial resistance.

Below, upon example of Quadrupole for E-989 [3] we are considering one possibility to feed the HV electrostatic elements with *mechanical* HV switching relays- Kilovacs and gigavacs.

## 2. KILOVAC AND GIGAVAC FAMILY

One type of HV relay is so called Kilovacs (see Addendum). Basically this is a mechanical relay with contacts moved by an electromagnet. Kilovac family interesting for our purposes includes K70A, K70B and K70C. The last one has three contacts which allow the central one switching between two positions. These Kilovacs are filled with pressurized $SF_6$ gas. Other Kilovacs are H-19 and H-17 are vacuumed with Tungsten electrodes, see Addendum.

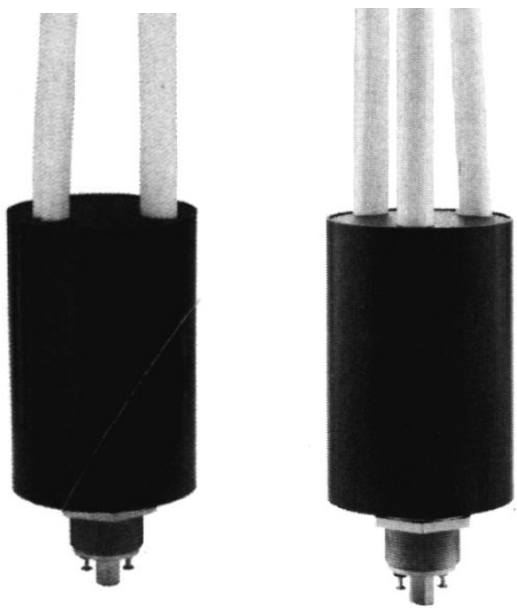

**Figure 1.** Model K70A,B at the left, K70C at the right. Diameter of the body is 2 inch. These Kilovacs commutate 70 *kV* at ~10 *A*.

Two parameters which characterize the switching time (see Addendum) are the operate (closing) and release times. The first parameter (closing time) is the time required for Kilovac to establish electric contact after a step of feeding voltage applied to the coil of electromagnet. Release time is a time between the moment, when the feeding voltage is turned off and the moment, when the electrical contact is broken. These times are indicated as they are of the order of 10-20 *ms*. These times associated mostly with *mechanical* motion of contacts from the initial position to the final one and do not reflect physical closing/opening times.

## 3. EXPERIMENT WITH KILOVAC

As we already mentioned, in a specification (see Addendum) for Kilovacs, the closing time indicated as long as 6 to 20 milliseconds, depending on model. The relese time is 15 *ms* maximun (for K70 family). This closing time indicates however the time counted from the beginnig of voltage pulse, which is feeding the magnet of kilovac, to the moment of *physical* closing of contacs. But for our purposes namely that last duration of electrical closing is important, as the time required for mechanical acceleration of contacs to the moment of physical closing of contacts is irrelevant for the closing time itself. One can think about this as of some delay; what is important- is the stability of this delay time from pulse to pulse.

Other specifics of *g-2* experiment E-989 taken as example, is that the the pulses are grouped in four trains by four pulses (16 total) [3]. Each pulse in a train separated by ~10 ms, the trains of four pulses each are separated by ~ 100 *ms* followed by deay time~920*ms*, so the period of sequence of all 16 pulses comes to 1.33 seconds, (i.e. 12 *Hz* for average power consumption balance). So there is a desire to turn off the quads, when there is no beam in a ring for reduction of sparking probability (which is proportional to the time when HV applied). So, definetely, there will be not a problem to turn of Kilovac for the interval of ~920 ms (time between trains containing 16 pulses) and for the time between train of four pulses (~ 100 ms).

To found what is the real closing time we used a simple arrangement with Kilvac H-12 in hand, Fig.2. This Kilovac has central contact and two others-one is "Normaly Cloced" to the central one and the other one is "Normally Open" to the central one. When the voltage applied to the kilovac coil, the central contact jumps to the one normally open.

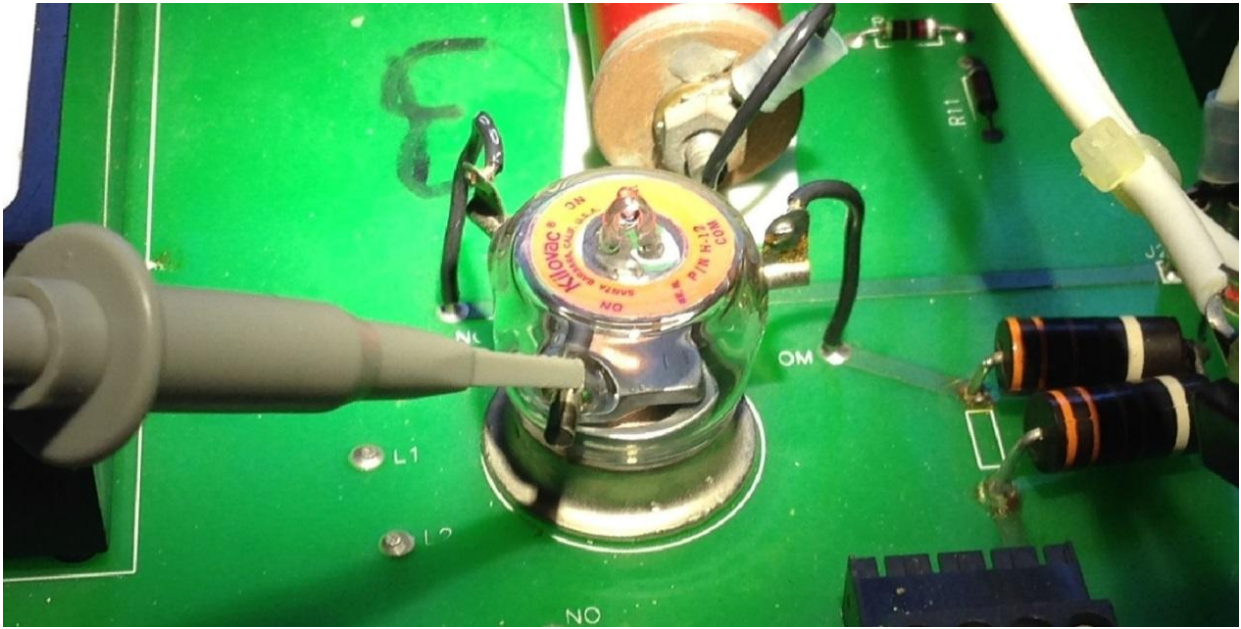

**Figure 2.** Setup for test switching time of Kilovac H-12 (12 *kVmax*, 30 *Amin,* cost~$650). The probe which is attached to the normally open contact has 1:10 attenuation.

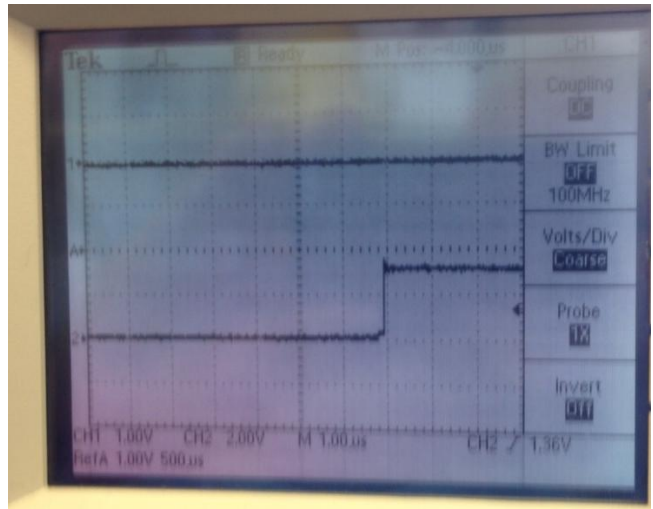

**Figure 3.** The rise time of voltage; horizontal time scale 1µs/div, lower ray. Driving voltage applied to the coil of Kilovac ~24 V.

One can see that the real contact time is well below a microsecond. To investigate dependence of opening/closing time on the feeding voltage applied to the coil, a dedicated voltage source was used able to work with regulating output voltage. It was found that minimal voltage required for Kilovac H-12 to close contact is ~12V. In Fig. 4 there is shown the reaction of Kilovac on applied voltage~ 15V.

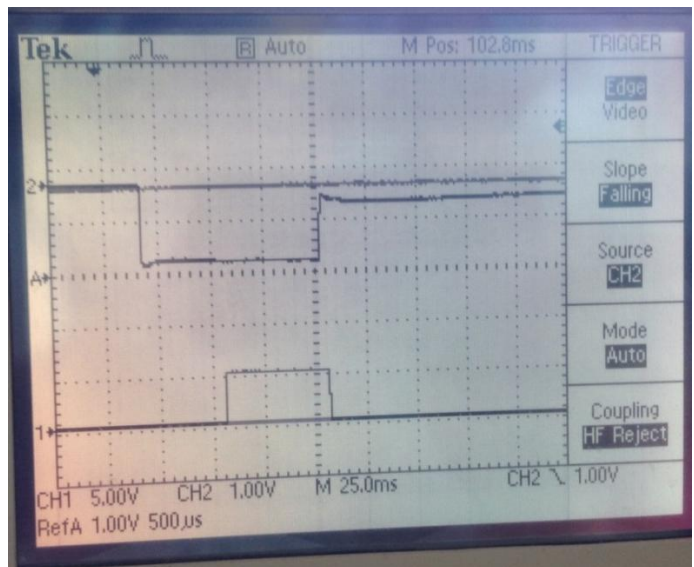

**Figure 4.** Upper ray-the driving voltage applied to the coil of Kilovac (negative, attenuated 1:10) Lower ray-is the voltage at Kilovac's "Normally Open" contact (positive, aqttanuated 1:10). Commutating voltage ~50V; 25 *ms/div*.

One can see that the closing time delay comes to ~ 30 ms, and the opening time delay comes to ~ 7ms. So there is a possibility for optimization of operation by changing the time profile of the voltage applied to the coil of electromagnet. It looks that the profile of voltage close to the triangle (trapezoidal, more exactly) might be the best one.

## 4. SCHEME WITH KILOVACS FOR QUADRUPOLES OF *g-2* EXPERIMENT

So as the HV Quadrupole (Dipole etc.) does not required current for its steady operation, just for the moments of charging/recharging, the thyratrons could be excluded. The charging current could be restricted to the value handled by Kilovac by choosing appropriated resistors in series. So for the limiting current, say 50*A* and voltage 30*kV*, the resistor value should be not less, than $R=30kV/50A=600\Omega$ only. To keep the resistor value located at the input of Quad higher than this minimal value, allows limiting the discharge current in a case of sparking between the electrodes and ground.

An example of quadrupole PS for *g-2* experiment is represented in Fig. 4. High frequency transformer, rectifying diodes, Kilovacs, resistors are located at the input of HV of the ring vacuum chamber in a sealed canister (see Fig.6). All elements of PS are immersed in Fluorinert for better insulation and cooling. Dielectric permittivity of Fluorinert $\varepsilon\sim1.9$ does not add much to the parasitic capacitance.

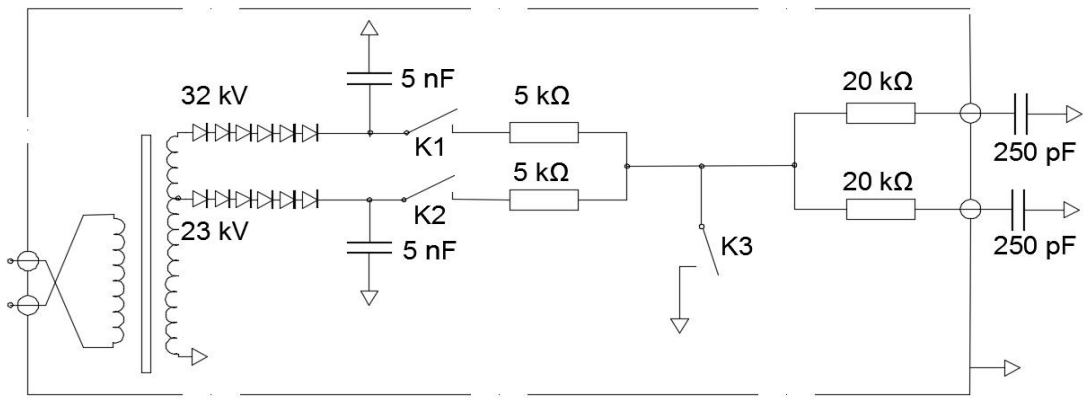

**Figure 4.** An example of electrical scheme of the Quadrupole PS for the positive polarity. For negative polarity the diodes with reversed polarity uses the same transformer taps. Kilovacs marked by K1-K3; for K1-K3 the K70A used, for K3 the model K70C could be used alsoused. Entire volume filled with Flourinert. The scheme is shown in a neutral position.

Simultaneous closing of K1 and K2 in a scheme in Fig.4 does not create a problem as the loop from lower HV will be electrically closed. So the total number of Kilovacs required with this scheme per quad is three for polarity which HV value should be changed[1] comes to five totals. Transformer is feed by high frequency bi-polar pulses and operates at few tens kHz (50-70*kHz*). This allows transformer to be a compact unit. There are lots of high voltage diodes operating at this frequency. Semtech© SHVM15F may serve as example; this diode holds 15*kV* reversed voltage and direct DC current is limited by 0.35A while repetitive pulsed one could be as high as 8*A*. Feeding of Kilovacs are going from special board with transistors amplifiers delivering 1*ms*-long pulses of 12-24*V*.

---

[1] Other polarity does not require change of voltage during entire presence of the beam in a ring .

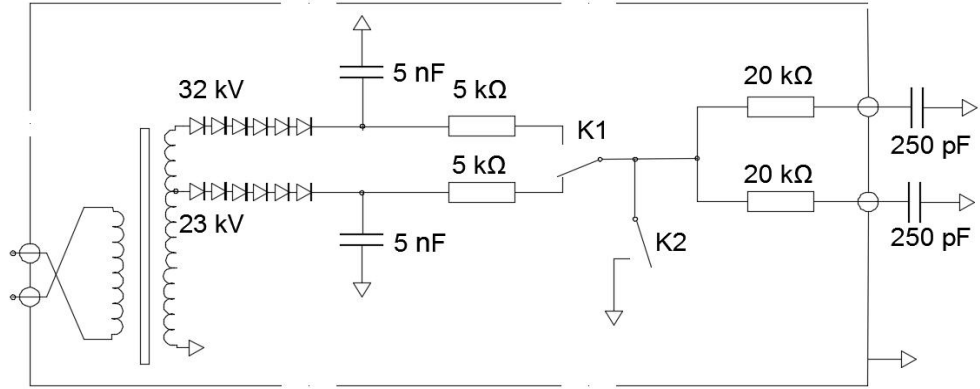

**Figure 5.** The scheme with two Kilovacs. This scheme is useful for switching between two voltages when usage of K2 is occasional. For K1 the K70C should be used.

The scheme in Fig. 5 might be useful if the switching between voltages could be of the order of 10-20*ms*. It might be useful for the (*g*-2) E-989 time pattern also, if the 23 kV voltages turned on in advance, and switching to the 32 *kV* happen ~15 µs after injection. Voltages and timing should be arranged so, that grounding of electrodes with K2 happens while the contacts of K1 are traveling between two states (from 32*kV* to 23 *kV*).

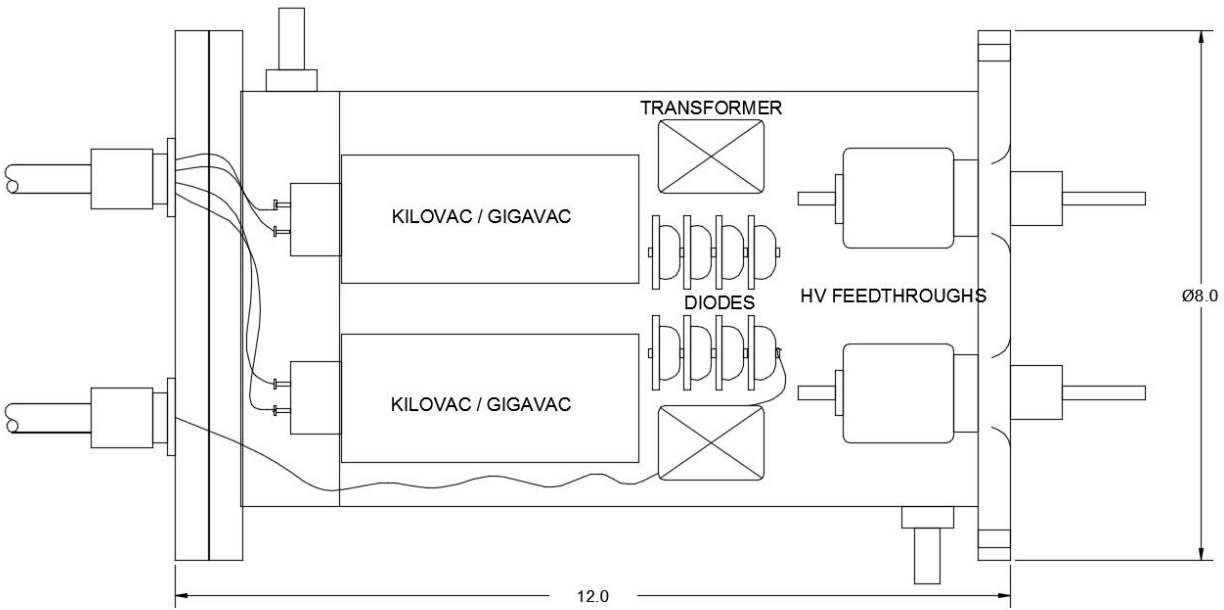

**Figure 6.** The PS schematics from Fig.4 and Fig.5. Dimensions are given in inches (Ø8", height = 12"); this PS supply mounted on the input flange of HV Quadrupole input. Entire volume filled with Fluorinert.

The transistor board for switching Kilovacs is not shown here; it is located in close vicinity.

## DISCUSSION

Usage of Kilovacs and Gigavacs in HV accelerator elements such as electrostatic Dipoles, Quadrupoles, etc. and for charging capacitors, open an interesting possibility for simplification of these devices and lowering theirs cost. Implementation of Kilovac in each case should be done after test of such device at full voltage and frequency of switching rate (100 *Hz* at max, 12 *Hz* average for *g-2*).

Short time of physical contact closing demonstrated is not surprising, as the sparking devices are widely used in a commutation elements; one can think, that in a case of mechanical relays, the triggering happens when the contacts are close enough, while in switches the triggering arranged with the help of initiating discharge or arrangement of instant overvoltage of the gap (for three-electrode sparking switches). As far as the release time, it should be investigated for each specific model. The possibilities here are in lowering and profiling feeding voltage and current. More radical could be usage of *magnetized* core of electromagnet, so it becomes sensitive to polarity of the feeding current. The relays with latching contacts operated with two separate coils for closing and opening open wide possibilities for optimization of open/closing times. One example of such devices are G64L and G71L from Gigavac®, see Addendum.

**ADDENDUM.**
*Screenshots from official sites of Companies*

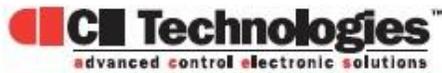

### Kilovac H-19  Make & Break Load Switching

**20 / 25 kV**

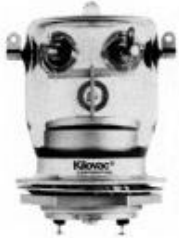

**Features:**
- 20 kV operating voltage
- Vacuum dielectric and tungsten contacts for power switching low current loads
- Double pole, double throw contacts
- Available with corona shield connectors
- Meets requirements of MIL-R-83725

### Kilovac H-17  Make & Break Load Switching

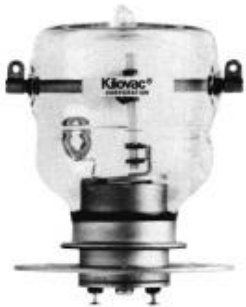

**Features:**
- Will isolate 12 kV at 32 MHz
- Tungsten contacts suitable for power switching low current loads
- Available with corona shield connectors
- Meets requirements of MIL-R-83725
- QPL version available, M83725/2

### PRODUCT SPECIFICATIONS

| Part Number | Units | H-19 | H-17 |
|---|---|---|---|
| Contact Arrangement | | DPDT | SPDT |
| Contact Form | | 2C | C |
| Test Voltage (dc or 60Hz) | kV Peak | 25 | 30 |
| Rated Operating Voltage | kV Peak | | |
| dc or 60 Hz | | 20 | 25 |
| 2.5 MHz | | 15 | 20 |
| 16 MHz | | 10 | 15 |
| 32 MHz | | 7 | 12 |
| Continuous Carry Current, Maximum | Amps | | |
| dc or 60 Hz | | 30 | 30 |
| 2.5 MHz | | 18 | 16 |
| 16 MHz | | 9 | 10 |
| 32 MHz | | 6 | 8 |
| Coil Hi-Pot (V RMS, 60 Hz) | | 500 | 500 |
| Contact Capacitance | pF | | |
| Between Open Contacts | | 1 | 1 |
| Open Contacts to Ground | | 2.5 | 2.5 |
| Contact Resistance, Maximum | ohms | 0.015 | 0.015 |
| Operate Time, Maximum | ms | 30 | 25 |
| Release Time, Maximum | ms | 20 | 25 |
| Shock, 11 ms 1/2 Sine | Peak G's | 30 | 20 |
| Vibration, 10 G's Peak | Hz | 55-500 | 55-500 |
| Operating Ambient Temperature Range | °C | -55 to +125 | -55 to +125 |
| Mechanical Life (Operations x 10⁶) | Cycles | 1 | 1 |
| Weight, Nominal | oz. | 8.5 | 7 |

### COIL DATA

| Nominal, Volts dc | | 12 | 26.5 | 115 |
|---|---|---|---|---|
| Pickup, Volts dc, Maximum | | 8 | 16 | 80 |
| Drop-Out, Volts dc | | .5 - 5 | 1 - 10 | 5 - 50 |
| Coil Resistance (Ohms ±10%) | H-19 | 48 | 225 | 2100 |
| | H-17 | 24 | 120 | 2900 |

Ratings listed are for 25°C, sea level conditions

### PART NUMBER SELECTION

Sample Part No. H- `17` `/12Vdc`
Model
 H-19
 H-17
Coil Voltage
 Blank = 26.5 Vdc
 /12Vdc = 12 Vdc
 /115Vdc = 115 Vdc

CII Technologies / P.O. Box 4422 / Santa Barbara, CA 93140 / ph 805.684.4560 / fax 805.684.9679 / e-mail info@kilovac.com / website www.ciitech.com   50

# CII Technologies
### advanced control electronic solutions

# 70 kV

## Kilovac K70A, K70B  *Make Only*

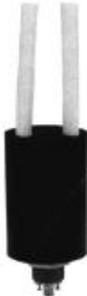

**Features:**
- New, small, compact 70 kV relay package
- SF-6 gas-filled for capacitive discharge and high voltage isolation applications
- Ideal for charging and discharging of high voltage capacitors
- Safe for use in adverse environments

## Kilovac K70C  *Make Only*

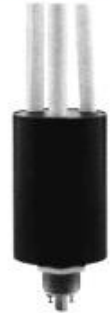

**Features:**
- SPDT version of K70A
- SF-6 gas-filled for capacitive discharge and high voltage isolation applications
- Ideal for charging and discharging of high voltage capacitors

COST ~ $1400.

| PRODUCT SPECIFICATIONS | | | | |
|---|---|---|---|---|
| Part Number | Units | K70A | K70B | K70C |
| Contact Arrangement | | SPST-NO | SPST-NC | SPDT |
| Contact Form | | A | B | C |
| Test Voltage (dc) | | 75 | 75 | 75 |
| Rated Operating Voltage | kV Peak | | | |
| dc | | 70 | 70 | 70 |
| 60 Hz RMS, kV Peak | | 30 | 30 | 30 |
| 2.5 MHz | | - | - | - |
| 16 MHz | | - | - | - |
| 32 MHz | | - | - | - |
| Continuous Carry Current, Maximum | Amps | | | |
| dc or 60 Hz | | 10 | 10 | 10 |
| 2.5 MHz | | - | - | - |
| 16 MHz | | - | - | - |
| 32 MHz | | - | - | - |
| Coil Hi-Pot (V RMS, 60 Hz) | | 500 | 500 | 500 |
| Contact Capacitance | pF | | | |
| Between Open Contacts | | - | - | - |
| Open Contacts to Ground | | - | - | - |
| Contact Resistance, Maximum | ohms | 2.0* | 2.0* | 2.0* |
| Operate Time, Maximum | ms | 20 | 20 | 20 |
| Release Time, Maximum | ms | 15 | 15 | 15 |
| Shock, 11 ms 1/2 Sine | Peak G's | 20 | 20 | 20 |
| Vibration, 10 G's Peak | Hz | 55-500 | 55-500 | 55-500 |
| Operating Ambient Temperature Range | °C | 0 to +85 | 0 to +85 | 0 to +85 |
| Mechanical Life (Operations x $10^6$) | Cycles | 0.5 | 0.5 | 0.5 |
| Weight, Nominal | oz. | 18 | 18 | 18 |

| COIL DATA | |
|---|---|
| Nominal, Volts dc | 26.5 |
| Pickup, Volts dc, Maximum | 22 |
| Drop-Out, Volts dc | 1 - 10 |
| Coil Resistance (Ohms ±10%) | 75 |

Ratings listed are for 25°C, sea level conditions

### PART NUMBER SELECTION

Sample Part Number K70 [A] [8] [4] [1]

Contact Form
- A = SPST-NO
- B = SPST-NC
- C = SPDT

Coil Voltage
- 8 = 26.5 Vdc, Turret Terminal

High Voltage Connections
- 4 = Flying Leads, 12"
- 7 = Flying Leads, 72"
- 8 = Flying Leads, 36"

Mounting
- 1 = Threaded

* Contact resistance for gas-filled relays measured at 26 Vdc, 1 Amp

CII Technologies / P.O. Box 4422 / Santa Barbara, CA 93140 / ph 805.684.4560 / fax 805.684.9679 / e-mail info@kilovac.com / website www.ciitech.com

## G71L - LATCHING
**Make Only Load Switching**
**RoHS Compliant**

**70 kV**

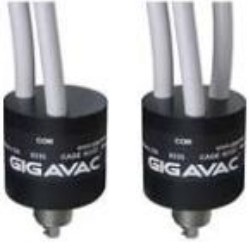

### FEATURES
- Latching coils for low power consumption and to ensure relay will remain in last position, even when no voltage is applied to the coil

**G71LA**
- Sealed 70 kV isolation in about 6.3 cubic inch (100 cubic cm) package - your will be system the smallest in the industry!
- SF-6 gas filled for capacitive discharge with no x-ray emissions
- Pre-tinned, optional length, high voltage leads making installation easy fast and easy
- Great for charging & discharging capacitors with no bounce under load
- GIGAVAC exclusive replaceable coils

**G71LC**
- Same great features as the G71A and G71B
- Double Throw (SPDT) for polarity reversal and multiple power source switching

### PRODUCT SPECIFICATIONS

| Contact & Relay Ratings | Units | G71LA | G71LC |
|---|---|---|---|
| Contact Form | | A | C |
| Contact Arrangement | | SPST | SPDT |
| Voltage, Test Max., Contacts & to Base (15 µA Leakage Max, dc or 60Hz) | kV Peak | 75 | 75 |
| Voltage, Operating Max., Contacts & to Base (15 µA Leakage Max.) | | | |
| dc | kV Peak | 70 | 70 |
| 60 Hz RMS | kV Peak | 30 | 30 |
| 2.5 MHz | kV Peak | - | - |
| 16 MHz | kV Peak | - | - |
| 32 MHz | kV Peak | - | - |
| Current, Continuous Carry Max | | | |
| dc or 60 Hz | Amps | 10 | 10 |
| 2.5 MHz | Amps | - | - |
| 16 MHz | Amps | - | - |
| 32 MHz | Amps | - | - |
| Coil Hi-Pot (V RMS, 60 Hz) | V | 500 | 500 |
| Capacitance | | | |
| Across Open Contacts | pF | - | - |
| Contacts to Ground | pF | - | - |
| Resistance, Contact Max @ 1A, 28 Vdc | ohms | 2.0* | 2.0* |
| Operate Time | ms | 20 | 20 |
| Reset Time | ms | 15 | 15 |
| Life, Mechanical | cycles | .5 million | .5 million |
| Weight, Nominal | g (oz) | 336 (12) | 336 (12) |
| Vibration, Operating, Sine (55-500 Hz Peak) | G's | 10 | 10 |
| Shock, Operating, 1/2 Sine 11ms (Peak) | G's | 20 | 20 |
| Temperature Ambient Operating | °C | -50 to +85 | -50 to +85 |

### COIL RATINGS

| | |
|---|---|
| Nominal, Volts dc | 26.5 |
| Pick-up, Volts dc, Max. | 22 |
| Reset, Volts dc | 1 - 10 |
| Coil Resistance (Ohms ±10%) | - |

Ratings listed are for 25°C, sea level conditions.
Coils are polarity sensitive.
Observe polarity marked on coil terminals.

For more information, refer to
**Relay User Instructions**

**G71   L   A   8   4   1**

**Latching Designator**

**Contact Form**
A = SPST
C = SPDT

**Coil Voltage ****
8 = 26.5 Vdc, Turret Terminal

**High Voltage Connections**
4 = Flying Leads, 12"
7 = Flying Leads, 72"
8 = Flying Leads, 36"

**Mounting**
1 = Threaded

*Contact resistance for gas-filled measured at 28 Vdc, 1 Amp
**Order the relay with the part number as shown. The latching "L" designator and the coil voltage will not appear in the P/N on the relay but will be indicated on the label that is on the base of the relay. Observe coil polarity.

09/06/11